# Far-field fluorescence microscopy beyond the diffraction limit: Fluorescence imaging with ultrahigh resolution

James H. Rice
School of Chemical Sciences and Pharmacy,
University of East Anglia, Norwich NR2 3RG, UK.

## Abstract

Fluorescence microscopy is an important and extensively utilised tool for imaging biological systems. However, the image resolution that can be obtained has a limit as defined through the laws of diffraction. Demand for improved resolution has stimulated research into developing methods to image beyond the diffraction limit based on far-field fluorescence microscopy techniques. Rapid progress is being made in this area of science with methods emerging that enable fluorescence imaging in the far-field to possess a resolution well beyond the diffraction limit. This review outlines developments in far-field fluorescence methods which enable ultrahigh resolution imaging and application of these techniques to biology. Future possible trends and directions in far-field fluorescence imaging with ultrahigh resolution are also outlined.

# Introduction

The term fluorescence was introduced by George Stokes in 1852 to describe the red light that was emitted from the mineral fluorspar when illuminated by an ultraviolet light source [1]. Following this it was found that a large number of materials fluoresce when irradiated with ultraviolet and visible radiation. In order to observe the fluorescence emission from materials on small length scales work was undertaken to utilise the technique of optical microscopy. This approach enabled Köhler et al in 1904 to build what is considered to be the first fluorescence microscope [2]. Developments in fluorescence microscopy were sufficient by the 1930's to enable imaging of fluorescent components in bacterial cells [3]. Today, the technique of fluorescence microscopy is an essential tool in biology and the biomedical sciences due to attributes that are not readily available in other methods of microscopy.

Fluorescence microscopy enables spatial selection of specific areas in a biological sample to be imaged by directing the radiation source to the specified region and then collecting the resulting fluorescence (as outlined schematically for a cell in Fig. 1). However, this ability to spatially select the area of study has a physical limit, the diffraction limit. Abbe at the end of the 19$^{th}$ century showed that the smallest distance that can be resolved between two lines using an optical microscope had a limit which is referred to commonly as the diffraction limit [4,5]. The diffraction limit is proportional to the wavelength and inversely proportional to the angular distribution of the light observed. This means that any optical microscope has a theoretical limit to its resolution. The resolution limit is often described by the minimal fluorescence spot size, which is calculated using the mathematical function referred to as the point-spread function (PSF) [6]. Since the diffraction limit originates from the fact that it is impossible to focus light to a spot smaller than approximately half the excitation wavelength. In practice this means that the maximal resolution (as expressed through the PSF) in fluorescence microscopy is c.a. 250 nm.

The demand in biology for improved resolution has stimulated research into creating fluorescence microscopy techniques that exceed the diffraction limit. A number of fluorescence microscopy techniques have demonstrated images with a resolution well beyond this limit. Different approaches to enable fluorescence microscopy to image with ultrahigh resolution beyond the diffraction limit have now emerged. These methods can be divided into near-field and far-field techniques. Near-field fluorescence microscopy is a powerful imaging technique that enables images to be recorded with resolution well beyond the diffraction limit by detecting fluorescence from the sample surface using a nanometre sized tip. However, in many applications in the life sciences collecting the light far away from the sample is required.

This has stimulated research into developing methods to image beyond the diffraction limit based on far-field fluorescence microscopy techniques. Rapid progress is being made in this area of science with methods enabling fluorescence images taken in the far-field to possess a resolution down to tens of nanometres. These developments in far-field fluorescence imaging with ultrahigh resolution are opening up a new area in optical instrumentation. This review centres on far-field fluorescence methods which enable ultrahigh resolution imaging, outlining developments in experimental techniques and applications of these techniques to biology. Future possible trends and directions in fluorescence imaging with ultrahigh resolution are also outlined.

# Far-field fluorescence imaging with ultrahigh resolution

Techniques based on far-field fluorescence microscopy have emerged that have successfully enabled ultrahigh resolution fluorescence imaging beyond the diffraction limit. The use of linear fluorescence methods and non-linear fluorescence methods based on such phenomena as standing waves and saturable transitions have been successfully utilized to enable ultrahigh resolution far-field fluorescence imaging. In addition, techniques using mathematical determination and advanced optics have also been applied to enable fluorescence imaging with ultrahigh resolution in the far-field.

## The use of linear processes

### Standing wave fluorescence microscopy

A fluorescence microscopy technique using standing waves was shown by Bailey et al in 1993 to enable sub-diffraction limited fluorescence imaging [7]. This method called standing wave fluorescence microscopy (SWFM) provides a method with high spatial resolution coupled with high imaging speed, based on wide-field imaging.

In SWFM excitation light is split into two and directed into the each of two objectives in a counter propagation alignment (as shown in Fig 2). The counter propagating beams interfere to create the standing wave field. The relative phase of each of the beams is controlled by changing the path length the light travels using a piezoelectric shift of the beam splitter which can shift the standing wave field with a resolution of tens of nanometres. In this way SWFM is able to image with an axial resolution down to 50 nm [7].

There are several challenges to be overcome using SWFM. One of the draw backs of this method is that it requires samples with a thickness on the order of the wavelength of the excitation light i.e. as thin as a single standing wave. This limits the range of samples that can be studied with this technique. A number of experimental challenges exist that limits the image quality that can be obtained using SWFM. Defocusing, aberrations and irregularities on surfaces all can cause deformations in the standing wavefront planes which create poor image quality and reduces the image resolution. However, methods have been developed to overcome some of these issues. It has been demonstrated that the use of polarisation and carefully chosen optics eliminates deformations in the standing wavefront planes which with image processing algorithms has enabled enhancement in image quantity [8,9].

SWFM has been applied successfully to a number of different biological systems. Bailey et al recorded sub-diffraction limited images of the living growth phase 3T3 fibroblasts containing a fluorescent analogue of smooth myosin II enabling the myosin aggregates to be more clearly resolved compared to traditional fluorescence microscopy [10]. Freimann et al used SWFM to study anti-body stained dry cells, examining the distribution of α-smooth muscle actin near a nucleus with a resolution superior to that achieved by conventional fluorescence microscopy [8]. The image resolution provided by SWFM was applied by Abraham to determine the thickness of lamellipodia of migrating fibroblasts in living cells [11]. The reported thicknesses was measured to be sub 200 nm. In addition to these studies of structural biology, studies of protein dynamics using SWFM has been performed. Davis and Bardeen used two-photon SWFM to study the diffusive motion of DNA-containing chromatin in live cells and isolated nuclei [10]. The diffusion processes studies centred on understanding the interactions of histone proteins with DNA which occurred on 100 nm length scales which was accessible when imaging using two-photon SWFM. These studies demonstrated that the ultrahigh image resolution achievable using SWFM enabled better understanding of fundamental questions related to structural biology and also to protein biology.

An interesting development in SWFM instrumentation was demonstrated by Peter So and colleagues who coupled SWFM with total internal-reflection fluorescence (TIRF) microscopy, producing a technique referred to as SW-TIRFM [12,13]. SW-TIRFM provides extended image resolution by using the property of a standing evanescent wave's effective wavelength [14]. The standing evanescent wave has an effective wavelength of $\lambda/2n \sin \theta$. The physical resolution in the evanescent region of the low-index material is now defined by the wavelength of light in the high-index material which is increased by n compared to traditional microscopy. Using this approach selective excitation of the sample under study is achieved with a spatial resolution of around 100 nm [15]. In addition, the technique generates low levels of background fluorescence and minimum exposure of samples to light above the evanescent excitation area. Experimentally, SW-TIRFM uses a single objective with a laser excitation source and CCD detection system as shown in Fig 2. This technique allows images to be collected rapidly with accumulation times well under 1s. SW-TIRFM was used by Gliko et al to determine the size of biological nanostructures in living cells with a spatial resolution of less than 100 nm [16].

**4Pi microscopy**

4Pi microscopy was proposed in 1992 by Hell and Stelzer as a method of improving image resolution beyond the diffraction limit [17]. This technique, called 4Pi in reference to $4\pi r^2$, uses a standing wave field obtained in the focus of a microscope via two overlapping collimated laser light sources to generate images with sub-diffraction limited resolution.

4Pi is a confocalization fluorescence technique. 4Pi images are accompanied by image aberrations (specifically, axial side lobes) created from the inability of the two aligned objectives to provide a full spherical solid angle. It has been found that these side lobes can be suppressed through the use of confocalization [17]. This approach when used in combination with multiphoton excitation reduces the presence of axial side lobes and provides a higher contrast of the interference pattern [18]. Combining confocalization and multiphoton excitation in this way is known as multiphoton 4Pi confocal microscopy. In addition it was found that the utilisation of specific imaging algorithms i.e. an optical transfer function has been shown to significantly enhance the spatial resolution for multiphoton 4Pi confocal microscopy images [19].

In 4Pi microscopy samples are required to be thin, on the order of 200 nm, for the light to effectively transverse the sample. The laser excitation source is split into two arms of equal path length with the resulting fluorescence directed to a photodetector. The sample is illuminated by focusing excitation light through both objective lenses in a coherent manner onto the same spot and or by coherently detecting the emerging fluorescence light through both lenses. The result of this is that constructive interference of the two focused light fields produces an excitation point below the diffraction limit. The technique requires the use of a confocal microscope and consequently the sample is scanned through the focal spot to build an image. This requires the use of a scanning stage and leads to long accumulation times. A closely related technique called I$^5$M microscopy using interfering illumination and detection performed in the wide-field has enabled faster data acquisition and up to a 50% increase in fluorescence detection [20,21]. However, the lack of confocal or multi photon-excitation results in high side lobes that are difficult to deal with through image processing. In addition the performance of I$^5$M deteriorates with the aperture angle a result of which is that I$^5$M is difficult to combine with water immersion lens technologies. However, images with a resolution of around 100 nm have been recorded in fixed cells using I$^5$M [20].

Multiphoton 4Pi confocal microscopy has been effectively applied to study a range of biological systems. This technique has been applied to structural biology by imaging F-actin fibers in mouse fibroblast which enabled the diameter of the fibres to be determined to a (sub-diffraction limited) accuracy of 200 nm [22]. Schrader et al demonstrated that multiphoton 4Pi confocal microscopy could be applied to produce 3D images with sub-diffraction resolution [23]. 3D images were constructed of labeled actin filaments within a fibroblast cell by recording a stack of consecutive images. This method was applied to create 3D images of the microtubule network in a fixed mouse fibroblast cell with a resolution of up to 100 nm [24]. These studies contributed to fundamental questions concerning the understanding of structure in cellular systems. However, these studies were made using a glycerol based environment in which the biological material was mounted in order to aid the collection of fluorescence signal. Bahlmann and Hell showed that multiphoton 4Pi confocal microscopy could also be applied to aqueous environments [25]. This opened the way for the study of live cells to be imaged using 4Pi which was achieved by Bahlmann et al who applied multiphoton 4Pi-confocal microscopy to study membrane stained live Escherichia coli bacteria [26].

**Multi focal multi photon 4Pi-confocal microscope**

In order to create a faster image accumulation time multiple beam scanning has been applied to 4Pi. This approach is known as beam scanning multi focal multi photon 4Pi-confocal microscope (MMM-4Pi). Egner et al used up to 64 4Pi foci to scan across a sample to obtain an image with a 100 nm resolution in all three spatial axes [27]. Experimentally, MMM-4Pi is achieved by deflecting an array of 4Pi foci across the specimen with a tilting mirror (galvo) as shown in Fig. 2. The result of which is up to a 15 fold reduction in image acquisition time compared to 4Pi-confocal microscopy.

MMM-4Pi has been successfully applied to study 3D structure in live cells. 3D images of the mitochondrial compartment in yeast cells were obtained using MMM-4Pi with an image resolution of around 100 nm in all three spatial axes in both aqueous and glycerol media enabling the diameters of tubules in the mitochondrial reticulum to be measured with a precision of 150 nm [27]. Improvements have been made to the optical set-up of MMM-4Pi to provide 3D images with a spatial resolution of around 80 nm [28]. This set-up was applied to produce 3D sub-diffraction limited images of mitochondrial network of live Saccharomyces cerevisiae cells and Golgi-resident GFP fusion proteins of a live mammalian cell [27,29].

## The significance of non-linearity

The use of non-linear effects has attracted attention due to the potential in such effects to extend further image resolution. In particular a phenomenon called saturation has been utilised to extend the resolution of imaging in a number of fluorescence microscopy techniques.

**Stimulated emission depletion fluorescence microscopy**

Stimulated emission depletion (STED) fluorescence microscopy utilises the effect of saturated depletion of the fluorescent state of a fluorophore with a focal intensity distribution with a local zero-point, an approach proposed by Hell and Wichmann [30].

A STED microscope uses a duel pulse approach, an excitation pulse and a STED pulse. A critical feature of the set up is the utilisation of a spatial phase modulator to control the properties of the STED pulse. The STED beam has an altered transverse mode replicating a $TEM_{0i*}$ mode, sometimes referred to as the doughnut mode as the pulse possesses no intensity at the centre. The STED and excitation beams are overlapped and directed into a confocal microscope has outlined in Fig 2. This creates the conditions for stimulated emission depletion [31]. In a molecule, the absorption of a photon leads to the population of the fluorescence state $S_1$. This state may spontaneously decay to the ground state releasing a photon or through spin-orbital coupling, intersystem cross from the $S_1$ to a $T_1$ state. These processes occur in the absence of the STED pulse. The introduction of the STED beam creates the conditions for stimulated emission from the $S_1$ to the ground state. The distributed fluorophores that encounter the centre of the (doughnut shaped) STED pulse experience no STED intensity and relax spontaneously, however, flurophores that encounter positive STED intensity which occur at the edges of the donut shaped pulse relax via stimulated emission. Careful spatial and temporal alignment of the excitation and STED pluses and control over the transversal wave modal distribution can establish the criteria for an imaging with a resolution of under 100 nm [31].

The enhanced resolution of STED has been applied in several studies to resolve cellular structures under 100 nm in size. STED has been successful in advancing the understanding of neurotransmitter release in the synaptic vesicle membranes by imaging individual vesicles [32]. To achieve this Willig et al applied STED to image individual synaptic vesicles which were too small and densely packed for confocal fluorescence microscopy to resolve. Kittel et al applied STED to better understand synaptic communication by imaging the molecular organization of presynaptic active zones during calcium influx–triggered neurotransmitter release [33]. STED images enabled better understanding of the role of presynaptic Ca ion channels localization in vesicle release. Kellner et al imaged acetylcholine receptor supramolecular aggregates using STED [34]. The resolution provided by STED enabled the effect of methyl-β-cyclodextrin on the organization of acetylcholine receptor aggregates to be studied. Lin et al undertook STED studies of TRP channel M5 expressing olfactory sensory neurons using GFP expressed labels [35].

**T-Rex stimulated emission depletion fluorescence microscopy**

The resolution of STED can be further enhanced using a technique call T-Rex STED. This method enables an image resolution of 15–20 nm to be achieved [36]. The T-Rex STED technique uses the same experimental approach used in STED, but with a much reduced repetition rate.

One of the challenges to maximise image resolution using STED is photo-bleaching, with STED the more photo-bleaching is reduced the better the image resolution. In STED microscopy two bleaching routes are possible [36]. One of them involves the absorption of a photon by a molecule in the $S_1$ state leading to a $S_{1+n}$ state another is intersystem crossing of the $S_1$ state to the $T_1$ state. It is expected that the bleaching pathway leading to a $S_{1+n}$ state is counteracted by stimulated emission. Bleaching through internal conversion of the $S_1$-$T_1$ is very effective in STED as it occurs in a few picoseconds. It is expected that as long as the molecule remains in the singlet system, the STED pulse is able to protect the molecule from photo-bleaching. When the molecule is in the triplet state which is long lived, the molecule will be exposed to a series of intense pump and STED pulses. The absorption cross sections for the triplet state can be large and thus the STED and/or pump pulses can efficiently pump up the molecule to higher triplet states $T_{1+n}$. To avoid this bleaching mechanism requires minimum excitation when the molecule is in the triplet state. This can be achieved through the use of low laser pulse repetition rates which enables sufficient time for the molecules in the triplet state to return to the ground state. This experimental approach to STED is called triplet relaxation (T-Rex) STED.

T-Rex STED requires that the fluorophore possess a range of specific physical properties such as efficient spin-orbital interaction, photostability and a high absorption cross section. This withstanding T-Rex STED has been demonstrated using a number of different molecular flourophores e.g. Rhodamine dye Atto532, Dyomics485, Atto465, and carboxyl-Rhodamine 6G [36]. In addition, T-Res STED studies of a number of biosystems have been performed. Using labelled secondary antibodies with Atto532, synaptotagmin I transmembrane synaptic vesicle proteins were imaged revealing protein patches with an image resolution of up to 25 nm [36]. T-Rex STED was also applied to study the structure of mammalian interphase nuclei imaging the distribution of a speckle marker protein, SC35 with an Atto532-labeled antibody [36]. The resulting study produced images with resolution sufficient to resolve features which had only been seen previously using electron microscopy.

**Saturated structured-illumination microscopy**

Saturated structured-illumination microscopy (SSIM) is a technique that can theoretically create unlimited resolution in wide-field using a nonlinear fluorescence response and structured light. Using this approach Gustafsson demonstrated an image resolution of 50 nm using dye labelled beads [37].

SSIM centres on the generation of moiré fringes from carefully prepared structured illumination. Moiré fringes are produced when light at a set frequency intersects creating areas of increased intensity. In SSIM it is fluorescence centred moiré fringes that are generated. In order to achieve the best possible image resolution the excitation source should contain as high a set of spatial frequencies as possible. However, the preparation and observation of light with spatial frequencies is limited by diffraction. While this indicates that structured light will not lead to spatial enhancement beyond the diffraction limit, SSIM gets around this problem by utilising non-linear excitation processes in a fluorophore. Fringe patterns arising from non-linear emission from fluorophores will contain harmonic frequencies in addition to the fundamental moiré fringes which can result in an increase in the spatial resolution with a theoretically infinite resolution. Gustafsson showed that saturated structured-illumination could be achieved using a picosecond laser light source which was structured using a transmission phase grating and then passed into a wide-field microscope and the resulting fluorescence collected and detected with a CCD camera [37].

A requirement of this technique however is that the sample be motionless. This indicates that studies of biological systems using SSIM would be challenging, through theoretically possible. While SSIM has not been applied to image biosystems, it has been noted that structured-illumination fluorescence of biosystems has been demonstrated with images of actin cytoskeleton in fixed cells have been reported [38].

## Mathematically determined methods

Methods have been created using mathematical algorithms to interpret images obtained using diffraction limited florescence microscopy in order to enable spatial determination on the tens of nanometre scale and better. This approach centres on the determination of the PSF for individual flourophores.

**Fluorescence imaging with one-nanometer accuracy**

Fluorescence imaging with one nanometer accuracy (FIONA) uses images of single fluorophores which are diffraction limited [39-41]. Mathematical treatment of each fluorophore can be performed to determine their location with extremely high accurately i.e. 1.5 nm through the determination of the PSF [41]. Studies assessing various algorithms to accurately determine the position of the flourophores emission spot size have been made [40,41]. It has been demonstrated quantitatively that a two-dimensional Gaussian function creates the best fit to and in this way the position of the flourophore can be calculated with extremely high accuracy. Studies have shown that noise on an image induced by such phenomena as fluorescent background and camera readout can be minimized by collecting more photons when formulating an image. Using such as approach wide-field fluorescence microscopy studies of individual rhodamine dyes diffusing in a lipid bilayer and single GFP molecules in viscous solutions were tracked with 30 nm precision [42,43]. Studies using scanning confocal fluorescence microscopy of quantum dots have been able to resolve the distance between two nanocrystals to 6 nanometres [44].

FIONA has been applied to image molecular motors in combination with Total Internal Fluorescence Microscopy (TIFM). This approach has enables a spatial determination of 1.5 nm accuracy and has been applied to determine the mechanism of motion in molecular motors myosin V, kinesin, and myosin VI [41]. These studies were able to accurately determine the precise methods of motion in each molecular assembly by been able to measure average step sizes which were typical in the low tens of nanometres. However, imaging using FIONA has limitations. It has been reported that the orientation of the fluorophore molecule can induce a significant error when imaging with this method [45]. A single fluorophore molecule possesses a fixed emission dipole orientation and as a consequence its emission strongly depends on its 3-D orientation in space. This can introduce as much as 10 nm error in position determination.

**Point Accumulation for Imaging in Nanoscale Topography**

A method for sub-diffraction far-field fluorescence imaging based on the collisional flux of individual flourophores was demonstrated by Sharonov et al [46]. Point Accumulation for Imaging in Nanoscale Topography (PAINT) is based on monitoring solution diffusing individual flourophores interaction with the surface of a substrate. This approach is based on each probe that comes into contact with the substrate surface been immobilized and its location monitored with high precision by locating its centroid of its fluorescence by fitting a 2D Gaussian function (as with FIONA). The flourophore only generates a fluorescence signal when bound to the substrate surface. The specify nature of the binding was reported not to be significant. One of the key aspects to this approach is the requirement to keep the density of probes at the substrate surface dilute, i.e. below one molecule per micrometer squared during each observation period.

PAINT has enabled images of lipid bilayers to be recorded with a spatial resolution of 25 nm. These images resolved the contours of these bilayers, and large unilamellar vesicles 100 nm in diameter. PAINT allows rapid accumulation of images with high spatial resolution with the authors reporting micrometer-sized images can be recorded in seconds with a spatial resolution down of 25 nm. Each image was estimated to consist of large numbers of flourophores on the scale of $10^4$ per image. A number of flourophores were used to generate images demonstrating that a wide selection of probes may be applicable for PAINT imaging. Since labelling with flourophores is not required the PAINT method can be directly applied to image a range of systems without modifying the experimental approach. However, careful choice of flourophore is important to enable selectivity in imaging specific objects and systems, for example fluorescently labelled proteins are specific to particular cell compartments. To date PAINT has been used to image the surface of systems, however, it may also be possible to image the interior of systems also by depositing flourophores beneath the surface to the interior of systems.

**Photo-activated localization microscopy**

Betzig et al demonstrated a technique called photo-activated localization microscopy (PALM). PALM utilises photo-activatable fluorescent protein molecules to enable ultrahigh resolution images to be produced [47].

PALM utilises methods to resolve spatially the position of fluorophores within a diffraction limited spot by determining the fluorophores centre of fluorescence emission through a statistical fit (based on estimating the PSF). This is in common with both FIONA and PAINT, however PALM uses a method to isolate single molecules at high fluorophore densities based on the serial photoactivation and subsequent bleaching of a photoactivatable protein [47]. Experimentally PALM is performed using total internal reflection fluorescence (TIRF) microscopy to specifically target proteins in 50 to 80 nm thin sections and near the surfaces of fixed cells. This approach has reported to enable the localized of molecules within these sections with an accuracy of 10 nm.

In order to create a PALM image a series of images, each with individual fluorophores resolvable were recorded. These images were analysed and integrated together to form a single final PALM image. This was achieved by taking a consecutive series of images of the sample with an ensemble of fluorophores present in the image window. Following photo-bleaching and then photo-activation individual flurophores could be resolved and imaged. Betzig et al recorded a series of images consecutively over a time period of between 2 to 12 hours (with a single image been accumulation for c.a. 1s) in order to acquire a complete image stack that was used to create a single PALM image. Each PALM image was reported to consist of $10^5$ to $10^6$ fluorophore molecules. PALM experiments were performed on a series of fluorophores including green photoactivatable green fluorescent protein, yellow EKaede, Kikume Green-Red and Eos Fluorescent Protein [47]. PALM was used to study proteins bound to the plasma membrane, specifically the dEosFP-fused Gag protein of human immunodeficiency virus 1 [47].

## The use of advanced optics

Developments in optics are ever on going and have been applied to enable ultrahigh resolution far-field fluorescence imaging.

### Optical superlenses and hyperlenses

An approach using specially prepared lens systems referred to as superlens and hyperlens have been proposed and demonstrated to enable fluorescence imaging in the far-field with sub-diffraction limited resolution.

The core of this approach to imaging centres around recovering light waves that carry image information that are lost using traditional lens technology. During imaging information about structures that are smaller than the wavelength of the incident light (i.e. the excitation wavelength in the case of fluorescence) is carried in evanescent waves. The amplitude of these light waves decrease exponentially with distance. A consequence of this rapid loss in amplitude image resolution in the far field is lost.

The use of a hyperlens has been demonstrated to enable sub-diffraction limited fluorescence imaging in the far-field [48,49]. The use of a hyperlens possesses properties that enable evanescent fields to be turned into propagating waves enabling magnified far-field images of sub-wavelength structures. A hyperlens consists of a nanostructured metamaterial with a carefully designed shape that is cylinderic or half-cylinderic in geometry. This material has specifically prepared dielectric constants, specifically it consists of two dielectric materials that possess opposite values in two orthogonal directions [50,51]. This introduces anisotropy into the lens systems. A consequence of this is that limitations on propagating wavelengths that is characteristic of a conventional, isotropic medium is removed resulting in no diffraction limit, i.e. theoretically unlimited image resolution. In this way Liu et al used a curved, periodic stack of silver and aluminium oxide deposited on their half-cylinder cavity to enable sub-diffraction limited images with a resolution of 130 nm to be achieved [49].

An alternative lens technology that has been proposed to enhance image resolution in far-field imaging are superlenses [52]. Superlenses are made from a single layer of material with a negative refractive index. These lenses work on the principle that instead of converting evanescent waves into propagating waves, a superlens enhances the evanescent waves. A superlens enhances the evanescent waves by coupling to as surface plasmon polaritons that exist on the surface of the superlens. Smolyaninov et al made a superlens by creating concentric rings of polymethyl methacrylate (PMMA) on a gold surface [48]. The effect of these rings was to amplify the evanescent waves. Combining a hyperlens with a superlens was then undertaken to enable sub-diffraction limited fluorescence imaging in the far-field [48]. Liu et al developed an alternative superlens device by enhancing the conversion from evanescent to propagating waves through the presence on surface corrugations which enable stronger Plasmon resonances to be occur on the lens surface [53].

This approach to wide field imaging is very recent and has not been applied to image biosystems. To date the use of superlens and/or hyperlens technology has only been used to image lithography defined artificial systems such as nanowires in order to demonstrate sub-diffraction limited resolution. This approach offers a potentially inexpensive route for ultrahigh fluorescence imaging in the far-field.

## Discussion and future prospects

### The contribution to biology

Fluorescence imaging with ultrahigh resolution has enabled advances in the understanding of structure and processes in a number of biosystems to be made (see Tab. 1.).

The ultrahigh image resolution achievable using SWFM has enabled better understanding of fundamental questions related to structural biology and also to protein biology. A series of studies of live systems using SWFM enabled insights into cellular structure to be made. Studies of 3T3 fibroblasts using SWFM provided the increase in resolution above that of traditional fluorescence microscopy to enable myosin aggregate features to be resolved more clearly [7]. A further demonstration of the effectiveness of SWFM to advance the understanding of structural biology was made by determining the thickness of lamellipodia of migrating fibroblasts [11]. The reported thicknesses was measured to be under 200 nm a resolution better than that obtained with diffraction limited fluorescence microscopy. In addition to the understanding of structural biology advances in the understanding of protein dynamics has been made. SWFM has been applied to study the diffusive motion of DNA-containing chromatin in live cells and isolated nuclei [10]. This study centred on imaging processes involving histone proteins and DNA which occurred on 100 nm length scales something not possible using diffraction limited fluorescence microscopy.

At the same time as studies using SWFM were being performed the first set of studies applying 4Pi to biosystems were also being undertaken. A series of studies were performed which addressed fundamental questions relating to structural biology in stained and living systems through both 2D and 3D 4Pi imaging using a wide range of flourophores. MMM-4Pi imaging enabled 3D mapping of live eukaryotic cells with an image resolution of 100 nm [27]. This study enabled the morphology and the size of mitochondrial compartments to be determined. The effect of different growth media on the cellular structure was studied. The mitochondria of cells were shown to be sensitive to growth media as determined through the structure of the branched tubular reticulum. 3D sub-diffraction limited images of Golgi-resident GFP fusion proteins of a live mammalian cell were performed using 4Pi microscopy with an image resolution of around 100 nm [29]. This enabled detailed visualisation of the distribution of Golgi in all 3 spatial dimensions. This study was taken further through the application of dual-color 4Pi imaging of double stained Golgi stacks in a mammalian cell [54]. This enabled greater insight into the distribution of Golgi enzymes to be resolved by being able to discriminate between two enzyme types.

A series of recent studies have demonstrated that the ultrahigh image resolution achievable using STED can enable better understanding of neuroscience and structural biology. STED has been successful in advancing the understanding of neurotransmitter release in the synaptic vesicle membranes by imaging individual vesicles. Imaging of individual synaptic vesicles which were too small and densely packed for confocal fluorescence microscopy to resolve was achieved [32]. Better understanding of neuro-organisation was achieved by studying the TRP channel M5 expressing olfactory sensory neurons using GFP expressed labels as well as studying synaptotagmin I transmembrane synaptic vesicle proteins which revealed protein patches with an image resolution of 25 nm [35,36]. In addition to structural information, understanding of fundamental neuro-processes was also advanced using STED. Better understanding of synaptic communication was achieved by imaging the molecular organization of presynaptic active zones during calcium influx–triggered neurotransmitter release [33]. Better understanding of the role of presynaptic Ca ion channels localization in vesicle release was also achieved [34]. In addition to neuron based systems STED has been applied to image other biosystems. The structure of mammalian interphase nuclei was imaged showing the distribution of a speckle marker protein, SC35 with an Atto532-labeled antibody [36]. The resulting study produced images with resolution sufficient to resolve features which had only been seen previously using electron microscopy.

Studies of biosystems using SSIM, PAINT and PALM have been made but are confined to a single example for each imaging method. SSIM has successfully imaged actin cytoskeleton in fixed cells [38]. PAINT has imaged lipid bilayers with a spatial resolution of 25 nm which enabled contours of these bilayers, and large unilamellar vesicles 100 nm in diameter to be resolved [46]. PALM has imaged plasma membrane using the dEosFP-fused Gag protein of human immunodeficiency virus 1 [47].

It is noted that the ultrahigh resolution information that was obtained using the techniques mentioned above can be potentially obtained using other imaging techniques such as near-field and scanning probe methods. However, a number of significant advantages are attainable when imaging using far-field methods. Ultrahigh resolution far-field fluorescence techniques enable imaging of living systems, which are not possible using scanning probe methods such as TEM. Scanning probe methods are surface sensitive and are in general not applicable for imaging within living cells.

Optical near-field methods such as SNOM also possess limitations. The depth at which these methods can image is relatively limited. In addition 3D imaging is not feasible using near-field methods.

**Comparison of experimental approaches**

A comparison of each experimental approaches strengths and weakness has been outlined in Tab. 2. Arguably the simplest experimental systems are methods centring on the determination of the PSF for individual flourophores. No changes are required to standard fluorescence microscopes optical arrangements with the creation of images with ultrahigh resolution generated through mathematical treatment of diffraction limited images. However, work on understanding and improving mathematical methods to interpret diffraction limited images have shown that the positional determination performed by fitting a Gaussian to the emission intensity distribution of the fluorophore is open to error of up to 10 nm in determining spatial position induced through flux in the orientation of the fluorophores emission dipole. This withstanding excellent sub-diffraction limited resolution can be obtained. A key advantage of this approach is that it is a wide-field method enabling fast image accumulation times. PAINT uses this approach to generate micrometer-sized images which were reported to be recorded in seconds with a spatial resolution of c.a. 25 nm. However, careful choice of flourophore is important to enable selectivity in imaging specific objects and systems. While presently PAINT has been used to image the surfaces of biosystems it may also be possible to image the interior of systems by depositing flourophores to specific areas with the interior of the system under study. PALM in contrast requires a more involved experimental system based on a conventional TIRFM system and photo-active labels. In addition PALM requires long image accumulation times as long as 12 hours. Future developments in PALM are expected to arise from improvements in the properties of the fluorophore used. Enhanced photo-stability, efficient photo-activation and higher quantum yield of fluorescence will all contribute to better PALM images and shortened accumulation times. Improvements in mathematical algorithms will potentially enable more efficient extraction of data, reducing accumulation times and providing better image resolution.

Sub-diffraction limited imaging based on linear fluorescence measurements e.g. the experimental methods of SWFM, 4Pi and I$^5$M are challenging experimental methods. These experimental approaches require two very carefully aligned objectives with only ultra-thin samples being applicable for imaging. In addition these methods produce poor quality images which require the use of carefully chosen optics and image processing algorithms to enhance image quantity. However, work has been undertaken to improve experimental methodology. SWFM when used in combination with TIFR produces an improvement in image resolution by utilising the properties of evanescent waves. The resulting technique, SW-TIRFM, has an axial resolution of 100 nm and is a much simpler experimental method based on a single objective. However, the penetration depth of the evanescent wave limits the range of systems that are applicable for study. 4Pi has evolved a number of methods to improve image resolution, quality and image accumulation time. MMM-4Pi is arguably the most developed of the 4Pi techniques with the fast accumulation time, high spatial resolution and the use of confocal and multiphton excitation to enhance image quality. However, MMM-4Pi is a complex and expensive experimental approach. Recent developments in the use of single photon excitation and the use of cheap optics offers the opportunity that 4Pi can become experimentally simpler and cheaper. Lang et al who showed that one-photon fluorescence excitation, in place of multiphoton excitation, using novel semi-aperture optics can generate good quality images [54,55]. The utilisation of one-photon fluorescence in place of multi-photon excitation not only simplifies the experimental set-up for 4Pi but also reduces the image accumulation time as more fluorescence signal is generated.

The application of non-linear techniques STED and SSIM have demonstrated excellent image resolution. STED can be build using relatively inexpensive laser diodes, through these are required to match the fluorophore and as a consequence restrictions on the choice of fluorophore are present. In addition to this STED is a challenging experimental technique as preparing the beams with customised transverse wave properties is non-trivial. The image resolution of the STED technique is limited by the stability of the flourophores used. However, a range of flourophores have been successfully used to study a range of biosystems with image resolution under 100 nm. Improvements in the image resolution of STED has been demonstrated by using flourophores that phosphoresce rather than fluoresce, this approach being called T-Rex STED. The use of multi-colour imaging enables more detailed studies of biosystems. The challenge of multi-colour imaging of fluorophore sets in STED have began to be addressed. Donnert et al has recently demonstrated a new development in STED using a two-colour approach [56]. This approach uses two flourophores that are spectrally distinct. Specifically green and red emitting flourophores were used which on the basis of their wavelengths produced different PSF values, with a lateral resolution of 25 nm for the green and 65 nm the red fluorophores. This ultrahigh resolution spot size in combination with colocalizational understanding of the fluorphores opens up new opportunities for the understanding of biological systems. In contrast to STED (which has been applied to a number of biosystems), SSIM has only been applied to image a single biosystem (actin cytoskeletons). Studies on other biological systems would demonstrate a wider applicability of this experimental approach. SSIM uses relatively inexpensive optical components but the experiment itself is non-trivial with control over the wave patterns and image analysis challenging. However, this approach is attractive as it offers theoretically unlimited resolution.

The use of optical hyperlens and/or superlens technology has been shown to enable ultrahigh resolution fluorescence imaging in the far-field. This approach is attractive as it enables an inexpensive route for imaging with high resolution. Challenges in obtaining or making the lens are present as to date no commercial lenses of this type are available. However, potential exists for this approach to substantially contribute to the future development of ultrahigh resolution fluorescence imaging in the far-field.

**The role of the flourophore**

The choice of fluorophore is important in general for all fluorescence based techniques, and in particular for saturable absorption based techniques as the physical properties of the fluorophore are more demanding. Organic molecules have been used successfully in many studies but are prone to photo-bleach and are required to be chemically added to the biological system limiting their applicability to living cells. Proteins such as GFP are alternative labelling systems that can be used as genetically encoded markers. Protein mutagenesis offers a method to create flourophores with tailored optical properties reducing many of the problems associated with organic molecules. Quantum dots are another class of fluorophore that has been utilised. These materials are very photo-stable, possess a large stokes shift between their absorption and emission and can be made with a very wide range of wavelengths. Quantum dots are semiconductors made from metals such as cadmium and selenium and are poisons to living systems. In addition adding quantum dots to biosystems is non-trivial. However, quantum dots have been applied successfully to single tissue sections of biopsies for assessing malignant tumours and in 4Pi and STED microscopy [57,58]. Developments in fluorophore technology are being made in order to improve photostabillity and enhance fluorescence quantum yield. Widengren et al have demonstrated a method to chemically retard dye photobleaching and increase fluorescence yield [59]. Adding two types of anti-fading molecules, antioxidants and triplet-sate quenchers, were shown to increase the quantum yield of fluorescence and photostability of an organic dye molecule.

**The diffraction limit and optical microscopy**

Optical microscopy techniques based on multiphoton microscopy techniques such as second harmonic generation (SHG), sum frequency generation (SFG) microscopy and Coherent Anti-Stokes Raman spectroscopy (CARS) microscopy are increasing being applied to image biosystems. These techniques image resolution are limited by the diffraction limit to a spatial resolution on the order of the excitation wavelength. These techniques are attractive as they enable information to be obtained that is not attainable using fluorescence based methods. CARS microscopy is a nonlinear optical imaging technique in which pump and Stokes laser beams at different frequencies interact with a sample to produce an anti-Stokes (CARS) signal. This signal is strong when the frequency difference of the lasers corresponds to a Raman-active molecular vibration. CARS can produce images with strong contrast that chemically distinguish a specimen from its background. A key advantage with CARS is that no flourophores are required for imaging with this method. However, high laser intensities are required for imaging making photo-bleaching an issue. In addition the experimental set-up is non-trivial and requires expensive laser based equipment. Never-the-less this method has been applied successfully to image a number of biosystems [60,61]. Second Harmonic Generation (SHG) microscopy is a multiphoton technique that requires a polarizable material with non-centrosymmetric symmetry. SHG microscopy is a powerful method of investigating structures such as well-ordered protein assemblies including lipid membranes and muscle myosin [62, 63]. Using SHG it is possible to produce 2D or 3D images using a scanning confocal microscopy based technique [64]. Furthermore, SHG imaging using flourophores can also be performed which enable images to be recorded in systems with no well-ordered structure [65]. However, image accumulation times are long and images require significant deconvolution to recover image resolution. Photo-bleaching is a significant challenge as is the reduced number of molecular probes that are available for SHG compared to the number of flourophores available for fluorescence microscopy. Sum Frequency Gain (SFG) imaging is a non-linear optical process where a pulsed visible laser beam is overlapped with a tunable, pulsed IR laser beam to generate a signal at the sum frequency. SFG like SHG requires non-centrosymmetric symmetry [66]. The SFG provides a vibrational spectrum of the species under investigation. This method of imaging is prone to photo-bleaching and requires long accumulation times. In order to obtain the vibrational information images at a series of wavenumbers is required extending accumulation times. While all of these methods have been applied to image systems using near-field methods which have successfully enabled an image resolution well beyond the diffraction limit, to date, no advances have been made to image with a resolution using these techniques beyond the diffraction limit in the far-field.

The chemical characterisation ability provided through the vibrational sensitivity with methods such as SFG is attractive and provides in-depth understanding in regard to the chemical constituents of the material under investigation. However, such techniques are diffraction limited. However, advances in overcoming the diffraction limit in molecular vibration sensitive far-field optical techniques are attractive for imaging and are being pursued. An example of this is far-field infrared super-resolution microscopy which combines a fluorescence microscope and time-resolved transient fluorescence detected IR spectroscopy [67]. This technique enables IR imaging beyond the diffraction limit. The image resolution that can be obtained using this method is twice better than the diffraction limit for the IR wavelengths achieved through the use of visible laser excitation. This technique demonstrates that fluorescence based imaging can possess vibrational sensitivity and that imaging in this way can enable ultrahigh resolution. However, this approach is difficult experimentally to achieve and that the image resolution may be theoretically determined by the visible excitation wavelengths used, thus it will be challenging to significantly improve the image resolution using this experimental approach.

Optical near-field techniques have enabled fluorescence images with a resolution well beyond the diffraction limit to be obtained. Near-field scanning optical microscopy (NSOM) combines high-resolution scanning probe methods with fluorescence microscopy to enable ultrahigh resolution fluorescence imaging [68]. In NSOM the near-field optical probe is typically 20-120 nm in diameter. This aperture forms the resolution limit which is much smaller than the wavelength of the excitation light, thus providing sub-diffraction limited imaging. The generation of tips with such small apertures is demanding. The probe emits light that consists mostly of evanescent waves instead of propagating waves. As a result of this the probe can excite fluorophores only within a layer of <100 nm from the probe i.e. in the 'near-field' region. This limits significantly the studies that can be performed using SNOM, typically confining this method to surface related studies of biosystems. In addition, as it's a scanning probe based method using SNOM to study soft, rough and motile surfaces, such as the plasma membrane of living cells is very challenging [68]. However, SNOM has been applied to study a wide range of biological systems as is outlined in ref 68. The far-field ultrahigh resolution fluorescence methods that have been outlined above have comparable image resolution. Far-field based studies however enable studies of sub-surface structure for example been able to study internal cellular structure, something not possible at distances greater than 100 nm using SNOM. 3D imaging is also possible using far-field methods, something which is not possible using SNOM. Also the study of rough and soft surfaces is possible using far-field methods.

**Future directions**

Combining experimental methods is a powerful method to improve image resolution as has been shown for SWFM which when combined with TIRF microscopy to produce an enhanced microscopy technique SW-TIFRM as outlined above [12,13]. A technique that combines STED and 4Pi has been reported. It was shown that combining STED and 4Pi produces an enhancement in resolution and has been applied to image membrane-labelled bacteria and microtubules in fixed mammalian cells [69-71]. This was done in order to enhance image resolution. This approach however has been surpassed by developments in STED such as TRex-STED which produce better image resolution with a relatively simpler and inexpensive experimental set-up, which in addition does not require thin films or is constrained by demands for refractive index changes.

To fully understand biosystems the study of their dynamic processes is required. Wide field techniques such as SSIM and SWFM do allow for rapid image accumulation. Temporal resolution however is a challenge for confocal based techniques such as 4Pi and STED where scanning is required to create an image. A method based on line scanning has been shown to enable high-speed (100 frames per second) data imaging using confocal microscopy [72]. This approach can potentially lead to faster data acquisition in both 4Pi and STED techniques, opening up the possibility that these techniques may have accumulation times comparable to wide field methods.

It has been demonstrated that STED based imaging can allow information about the orientation of immobilized single molecules to be obtained [73]. This enables STED to be applied to orientation sensitive imaging in addition to spatial imaging of the flourophore. Dedecker et al outlined that based on the vector diffraction theory of light the use of doughnut modes can yield information about the orientation of immobilized single molecules allowing orientation sensitive imaging [73]. This approach potentially enables information about flourophore orientation as well as location to be extracted extending the information that can be obtained using STED. In addition to this, the authors proposed that the information in regard to orientation can be used to extend image resolution.

Studies at cryogenic temperatures offer a future direction in the development of ultrasensitive and ultrahigh resolution fluorescence microscopy. Low temperature measurements utilising the absence of phonon broadening leads to narrowing of the fluorescence emission wavelengths [74,75]. In addition increased fluorescence quantum yield and improved photostability are also expected. This will enable reduced image accumulation times and higher image quality. All of these factors are very attractive for ultrahigh resolution fluorescence imaging. While the use of cryogenic temperatures does not allow the study of living cells this approach is applicable for fixed cells.

Enderlein demonstrated far-field fluorescence microscopy technique that provided sub-diffraction limited imaging based on fluorescence decay [76]. The technique, referred to as dynamic saturation optical microscopy, uses the observed temporal dynamics of the fluorescence signal to provide information about the spatial distribution of fluorophores. This technique has enable images with a resolution for five times better than the diffraction limit. This approach can potentially be used in conjunction with other techniques to enable improvements in image resolution and offers interesting future directions in far-field fluorescence imaging.

**Implications for Abbe's rule**

The implications for linear and non-linear far-field fluorescence microscopy imaging techniques for Abbes rule was discussed by Stelzer [77]. It has been noted by Stelzer that images with sub-diffraction resolution obtained using the non-linear far-field fluorescence imaging techniques of STED and 4Pi could be interpreted not as images that possessed a resolution that broke the diffraction limit but as images that more accurately determined position. This line of argument does not extend to SWFM and structured illumination spectroscopy as these experimental approaches are based on the interference of light and do not break the diffraction limit as set by Abbe [5]. Photo-activated localization microscopy likewise does not break the diffraction limit as this technique utilizes mathematical analysis of individual fluorophores obtained at the diffraction limit in order to construct an image with sub-diffraction resolution. Hell in discussing the diffraction limit and far-field ultrahigh resolution fluorescence microscopy pointed out that breaking Abbe's rule requires being able to discern the number of flourophores that may be present from a densely packed feature that exists within a spatial distance covered by the diffraction limit [78]. Several methods have been able to discern numbers of flourophores present in such a densely packed feature. Achieved through, for example, sequentially recording this feature using "bright" and "dark" states to overcome diffraction limited image resolution. This approach is at the centre of techniques that use reversible saturable or photoswitchable transitions. On this bases Hell asserts that Abbes rule is broken.

As outlined above a number of methods using saturable or photoswitchable transitions have been applied in a number of studies of biological systems. The enhancement in resolution that this techniques bring have contributed to advancing the understanding of bioprocesses and biostrutures. These techniques together with those of interferance based methods and methods based on the use of advanced optics such as hyper lenses enable an image resolution far beyound that which would be estimated using Abbes rule. Today, Abbe's rule significance to microscopy is increasingly being marginalised.

## Conclusion

Far-field fluorescence microscopy is an important and extensively utilised tool for imaging biological systems. However, the image resolution that can be obtained has a limit as defined through the laws of diffraction. The demand in biology for improved resolution has stimulated research into developing methods to image beyond the diffraction limit based on far-field fluorescence microscopy techniques. This review has outlined notable emerging techniques that enable far-field fluorescence imaging with ultrahigh resolution. In addition the application of these techniques to biology and future possible trends and directions in far-field fluorescence imaging with ultrahigh resolution has also been outlined.

# References

1. G.G. Stokes, *J. Philos. Trans. Royal Soc. London*, **1853**, 143-149.
2. A. Köhler, *Physikalische Zeitschrift*, 1904, **5**, 666-673.
3. J.E. Barnard, F.K. Welch, *J. Roy. Micr. Soc*, 1936, **56**, 361-364.
4. E. Hecht, *Optics*, Addision and Wesley, 3$^{rd}$ edition, 1998, ch 10, 433-506.
5. E. Abbe, *Arch. Mikroskop. Anat.*, 1873, **9**, 413-420.
6. S.W. Hell, *Nature Nanotech.*, 2003, **21**, 1347-1355.
7. B. Bailey, D.L. Farkas, D.L. Taylor, F. Lanni, *Nature*, 1993, **366**, 44-48.
8. R. Freimann, S. Pentz, H. Holer, *J. Microscopy*, 1997, **187**, 193-200.
9. V. Krishnamurthi, B. Bailey, F. Lanni, *Proc. SPIE*, 1996, **2655**, 18-25.
10. S.K. Davis, C.J. Bardeen, *Biophys J.*, 2004, **86**, 555-564.
11. V.C. Abraham, V. Krishnamurthi, D.L. Taylor, F. Lanni, *Biophys. J.*, 1999, **77**, 1721-1732.
12. G.E. Cragg, P.T.C. So, *Optic Lett.*, 2000, **25**, 46-48.
13. P.T. So, H.S. Kwon, C.Y. Dong, *J. Opt. Soc. Am. A*, 2000, **18**, 2833.
14. H. Schneckenburger, *Curr. Opin. Biotechnol.*, 2005, **16**, 13.
15. E. Chung, D. Kim, P.T.C. So, *Optic Lett.*, 2006, **31**, 945-947.
16. O. Gliko, G.D. Reddy, B. Anvari, W.E. Brownell, P. Saggau, *J. Biomed. Optics*, 2006, **11**, 06097LRR.
17. S.W. Hell, E.H.K. Stelzer, *Optics Comm.*, 1992, 93, 277-282.
18. S.W. Hell, S. Lindek, E.H.K. Stelzer, *J. Mod Optics*, 1994, **41**, 675-681.
19. M. Gu, C.J.R. Sheppard, *Optics Comm.*, 1995, **114**, 45-49.
20. M.G.L. Gustafsson, D.A. Agard, J.W. Sedat, *J. Microscopy*, 1999, **195**, 10-16.
21. M.G.L. Gustafsson, *Curr., Opin., Struct., Bio.,* 1999, **9**, 627-634.
22. S.W. Hell, M. Schrader, H.T.M. VanderVoort, *J. Microscopy*, 1997, **187**, 1-7.
23. M. Schrader, K. Bahlmann, G. Giese, S.W. Hell, *Biophys. J.*, 1998, **75**, 1659-1668.
24. M. Nagorni, S.W. Hell, *J. Struct. Bio.*, 1998, **123**, 236-247.
25. K. Bahlmann, S.W. Hell, *Appl. Optics*, 2000, **39**, 1652-1658.
26. K. Bahlmann, S. Jakobs, S.W. Hell, *Ultramicro.*, 2001, **87**, 155-164.
27. A. Egner, S. Jakobs, S.W. Hell, *Proc. Natl. Acad. Sci. U.S.A.*, 2002, **99**, 3370–3375.
28. H. Gugel, J. Bewersdorf, S. Jakobs, J. Engelhardt, R. Storz, S.W. Hell, *Biophys. J.*, 2004, **87**, 4146–4152.
29. A. Egner, S. Verrier, A. Goroshkov, H.D. Soling, S.W. Hell, *J. Struct. Bio.*, 2004, **147**, 70-76.
30. S.W. Hell, J. Wichmann, *Optics Lett.*, 1994, **19**, 780-782.
31. K.I. Willig, R.R. Kellner, R. Medda, B. Hein, S. Jakobs, S.W. Hell, *Nature Meth.*, 2006, **3**, 721-723.
32. K.I. Willig, S.O. Rizzoli, V. Westphal, R. Jahn, S.W. Hell, *Nature*, 2006, **440**, 935-939.
33. R.J. Kittel, C. Wichmann, T.M. Rasse, W. Fouquet, M. Schmidt, A. Schmid, D.A. Wagh, C. Pawlu, R.R. Kellner, K.I. Willig, S.W. Hell, E. Buchner, M. Heckmann, S.J. Sigrist, *Science*, 2006, **312**, 1051-1054.
34. R.R. Kellner, C.J. Baier, K.I. Willig, S.W. Hell, F.J. Barrantes, *Neuroscience*, 2007, **144**, 135-143.
35. W. Lin, R. Margolskee, G. Donnert, S.W. Hell, D. Restrepo, *Proc. Natl. Acad. Sci. U.S.A.*, 2007, **104**, 2471-2476.

36. G. Donnert, J. Keller, R. Medda, M.A Andrei, S.O. Rizzoli, R. Luhrmann, R. Jahn, C. Eggeling, S.W. Hell, *Proc. Natl. Acad. Sci. U.S.A.,* 2006, **103**, 11440 -11445.
37. M.G.L. Gustafsson, *Proc. Natl. Acad. Sci. U.S.A.,* 2005, **102**, 13081-13086.
38. M.G.L. Gustafsson, *J. Microscopy*, 2000, **198**, 82-87.
39. M.K. Cheezum, W.F. Walker, W.H Guilford, *Biophys. J.,* 2001, **81**, 2378-2388.
40. R.E. Thompson, D.R. Larson, W.W. Webb, *Biophys. J.,* 2002, **82**, 2775-2783.
41. A. Yildiz, P.R. Selvin, *Acc. Chem. Res.,* 2005, **38**, 574-582.
42. H. Schindler, *Proc. Natl. Acad. Sci. U.S.A.*, 1996, **93**, 2926-2929.
43. U. Kubitscheck, O. Kuckmann, T. Kues, R. Peters. *Biophys. J.*, 2000, **78**, 2170-2179.
44. T.D. Lacoste, X. Michalet, F. Pinaud, D.S. Chemla, A.P. Alivisatos, S. Weiss, *Proc. Natl. Acad. Sci. U.S.A.*, 2000, **97**, 9461-9466.
45. J. Enderlein, E Toprak, P.R. Selvin, *Opt. Express*, 2006, **14**, 8116-8120.
46. A. Sharonov, R.M. Hochstrasser, *Proc. Natl. Acad. Sci. U.S.A.*, 2006, **103**, 18911–18916
47. E. Betzig, G.H. Patterson, R. Sougrat, W. Lindwasser, S. Olenych, J.S. Bonifacino, M.W. Davidson, J. Lippincott-Schwartz, H.F. Hess, *Science*, 2006, **313**, 1642-1645.
48. I.I. Smolyaninov, Y.J. Hung, C.C. Davis, *Science*, 2007, **315**, 1699–1701.
49. Z. Liu, H. Lee, Y. Xiong, C. Sun, X. Zhang, *Science*, 2007, **315**, 1686-1689.
50. Z. Jacob, L.V. Alekseyev, E. Narimanov, *Opt. Express*, 2006, **14**, 8247-8253.
51. A. Salandrino, N. Engheta, *Phys. Rev. B*, 2006, **74**, 075103-075109.
52. J.B. Pendry, Phys. Rev. Lett., 2000, 85, 3966-3969.
53. Z. Liu, S. Durant, H. Lee, Y. Pikus, N. Fang, Y. Xiong, C. Sun, X. Zhang, Nano Lett., 2007, 7, 403-408.
54. M.C. Lang, T. Müller, J. Engelhardt, S. W. Hell, *Opt. Express*, 2007, **15**, 2459 – 2467.
55. M.C. Lang, J. Engelhardt, S.W. Hell, *Optics Lett*., 2007, **32**, 259-261.
56. G. Donnert, J. Keller, C.A. Wurm, S.O. Rizzoli, V. Westphal, A. Schönle, R. Jahn, S. Jakobs, C. Eggeling, S.W. Hell, *Biophys J.*, 2007, **92**, L67-69L.
57. R. Medda, S. Jakobs, S.W. Hell, J. Bewersdorf J, *J. Struct. Bio.,* 2006, **156**, 517-523.
58. L.D. True, X.H. Gao, *J. Mol. Diagno. Science*, 2007, **9**, 7-11.
59. J. Widengren, A. Chmyrov, C. Eggeling, P.A. Lofdahl, C.A.M. Seidel, *J. Phys. Chem. A.,* 2007, **111**, 429-440.
60. T.B. Huff, J.X. Cheng, *J. Micro. Oxford*, 2007, **225**, 175-182.
61. X.L. Nan, A.M Tonary, A. Stolow, X.S. Xie, J.P. Pezacki, *ChemBioChem*, 2006, **7**, 1895-1897.
62. L. Moreaux, O. Sandre, S. Charpak, M. Blanchard-Desce, J. Mertz, *Biophys. J.*, 2001, **80**, 1568–1574.
63. T. Boulesteix, E. Beaurepaire, M.P. Sauviat, M.C. Schanne-Klein, *Opt. Lett.*, 2004, **29**, 2031–2033.
64. P.J Campagnola, A.C. Millard, M. Terasaki, P.E. Hoppe, C.J. Malone, W.A. Mohler, *Biophys J.*, 2002, **82**, 493–508.
65. A. Zoumi, X. Lu, S. Ghassan, J. Kassab, J. Tromberg, *Biophys. J.,* 2004, **87**, 2778–2786.
66. Y. Fu, H. Wang, R. Shi, J. Cheng, *Biophys. J.,* 2007, **92**, 3251-3259.
67. M. Sakai, Y. Kawashima, A. Takeda, T. Ohmori, M. Fujii, *Chem. Phys. Letts*., 2007, **439**, 171-176.

68. F. de Lange, A. Cambi, R. Huijbens, B. de Bakker, W. Rensen, M. Garcia-Parajo, N. van Hulst, C.G. Figdor, 2001, *J. Cell Sci.*, **114**, 4153-4160.
69. M. Dyba, J. Keller, S.W. Hell, *New J. Phys.*, 2005, **7**, 134-155.
70. M. Dyba, S.W. Hell, *Phys. Rev. Lett.,* 2002, **88**, 163901-164004.
71. M. Dyba, S. Jakobs, S.W. Hell, *Nature Biotechnol.*, 2002, **21**, 1303–4.
72. R. Wolleschensky, B. Zimmermann, M. Kempe, *J. Biom. Optics*, 2006, **11**, 05371RR
73. P. Dedecker, B. Muls, J. Hofkens, J. Enderlein, J.I. Hotta, Optics Exp., 2007, **15**, 3372-3383.
74. J.W. Robinson, J.H. Rice, R.A. Taylor, G.A.D. Briggs, R.A. Oliver, M.J. Kappers, C.J. Humphreys, *App. Phys. Lett.*, 2005, **86**, 213103.
75. J.H. Rice, J.W. Robinson, J.H. Na, R.A. Taylor, D.P. Williams, E.P. O'Reilly, A.D. Andreev, Y. Arakawa, S.I. Yasin, *Nanotechnology*, 2005, **16**, 1477.
76. J. Enderlein, *Appl. Phys. Lett.*, 2005, **87**, 094105-094108.
77. E.H.K. Stelzer, *Nature*, 2002, **417**, 806-807.
78. S.W. Hell, *Science*, 2007, **316**, 1153.

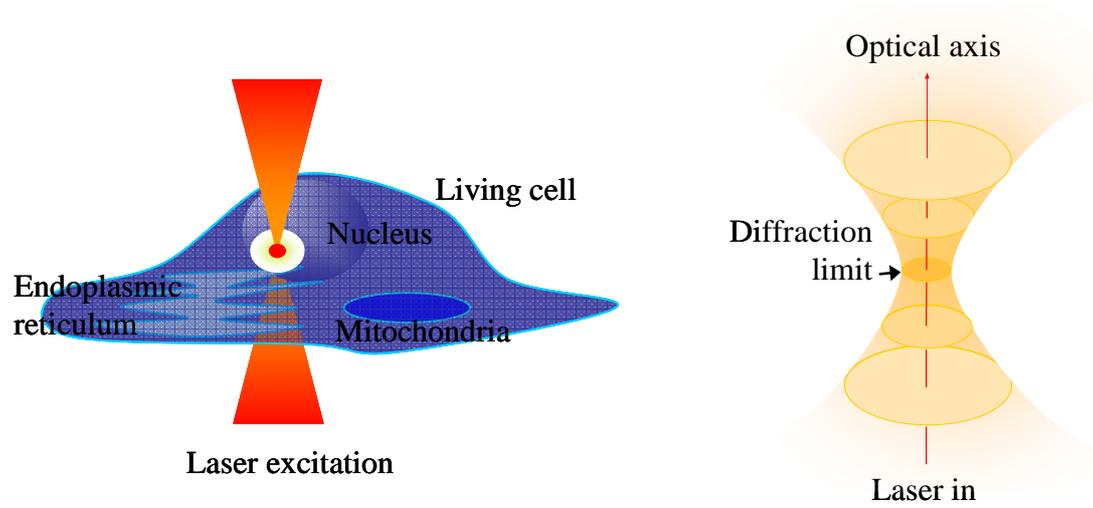

Fig 1. Schematic drawing outlining spatial selectively in an animal cell achievable with laser induced fluorescence microscopy, shown on the left. On the right, a schematic drawing outlining the diffraction limit spot size.

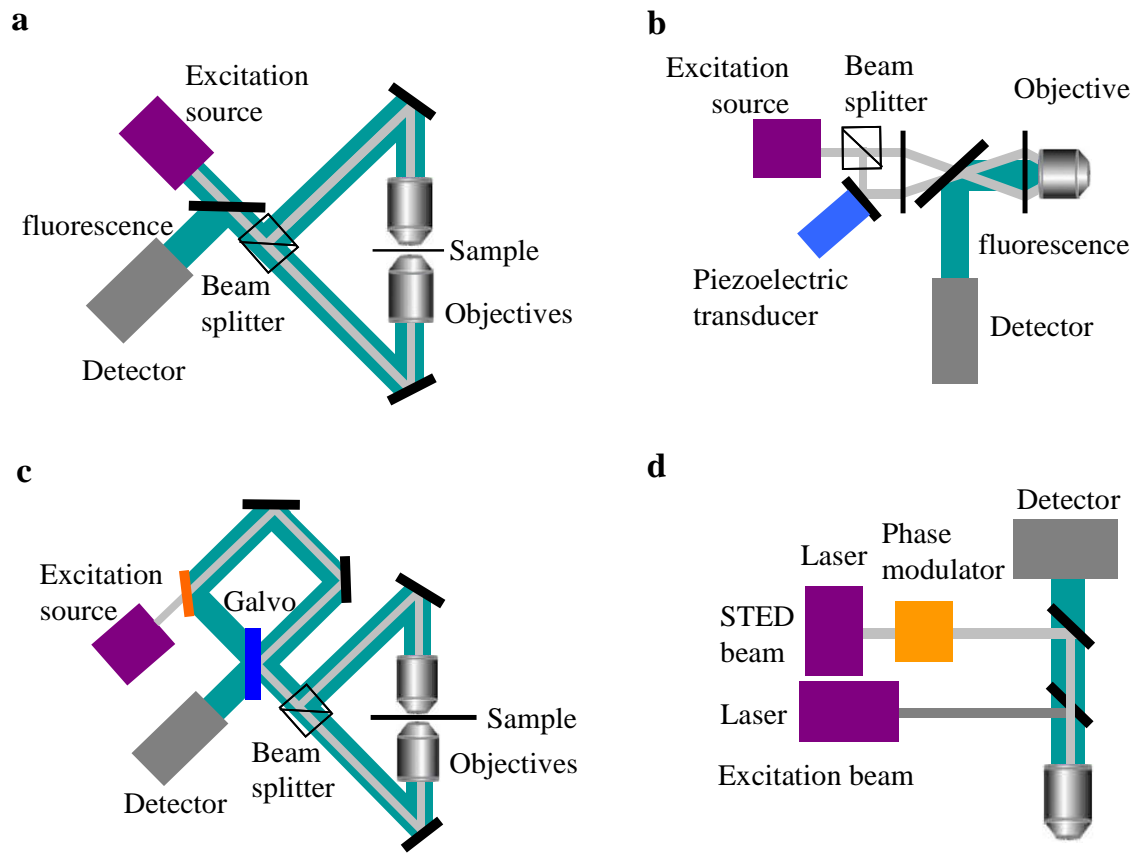

Fig 2. Schematic drawings of experimental set ups for (a) Standing wave fluorescence microscopy (SWFM), (b) combined standing wave fluorescence microscopy and total internal-reflection fluorescence microscopy (SW-TIRFM), (c) Parallelized multi photon 4Pi-confocal microscopy (MMM-4Pi), (d) stimulated emission depletion fluorescence microscopy (STED).

| Technique* | Biological system studied | Image Resol. (nm) | | Ref. No |
|---|---|---|---|---|
| | | Axial | Lat. | |
| **SWFM** | Living growth phase 3T3 fibroblasts containing a fluorescent analogue of smooth myosin II | 50 | - | 7 |
| | Distribution of α-smooth muscle actin in anti-body stained dry cells | 48 | - | 8 |
| | Diffusive motion of DNA-containing chromatin in live cells and isolated nuclei | 100 | - | 10 |
| | Thickness of the lamellipodia of migrating fibroblasts in living cells | 50 | - | 11 |
| | Biological nanostructures in living cells | 100 | - | 16 |
| **4Pi** | F-actin fibers in fixed mouse fibroblast cells stained using phalloidin-TRITC | 80 | 180 | 23 |
| | Immunolabelled microtubuli network in fixed NIH3T3 cells. | 190 | 190 | 26 |
| | GFP-labeled mitochondrial compartment of live Saccharomyces cerevisiae cells | 100 | 100 | 27 |
| | Mitochondrial network of live Saccharomyces cerevisiae cells labelled with GFP | 80 | 80 | 28 |
| | Golgi apparatus of a live mammalian cell labelled with GFP | 100 | 100 | 29 |
| | Microtubules in fixed PtK2 cells labelled with CdSe quantum dots and Oregon Green | 100 | 100 | 57 |
| **STED** | GFP-labeled rotavirus-derived particles and kidney epithelial PtK2 cell | 76 | 76 | 31 |
| | Individual Synaptic vesicle labelled with Atto 532-labelled antibodies | 66 | 66 | 32 |
| | Drosophila neuromuscular junction stained with monoclonal antibody Nc82 | 76 | 76 | 33 |
| | Acetylcholine receptor supramolecular aggregates stained with Alexa594-α-bungarotoxin | 71 | 71 | 34 |
| | TRPM5-expressing olfactory sensory neurons labelled with expressed GFP | 35 | 35 | 35 |
| | Synaptotagmin I, a transmembrane synaptic vesicle protein labelled with Atto532 | 15 | 15 | 36 |
| | Distribution of proteins SC35 in interphase nuclei labelled with an Atto532 antibody | 15 | 15 | 36 |
| **PAINT** | Distribution of unilamellar vesicles in lipid bilayers and a supported bilayer. | 25 | 25 | 28 |
| **PALM** | Lysosomal transmembrane protein CD63 tagged with tPA-FP Kaede. | 10 | 10 | 47 |

Tab. 1. List of ultrahigh resolution fluorescence imaging studies of biological systems. * SWFM is standing wave fluorescence microscopy, 4Pi microscopy, STED is stimulated emission depletion fluorescence microscopy, PAINT is point accumulation for imaging in nanoscale topography, PALM is Photo-activated localization microscopy.

| Technique* | Positive aspects | Negative aspects |
|---|---|---|
| **SWFM** | High spatial resolution down to 100 nm has been achieved. The technique enables high imaging speeds, based on wide-field imaging, with accumulation times well under 1s. | Ultrathin samples with a thickness on the order of the wavelength of the excitation light are required. Experimentally image quality is limited by defocusing, aberrations and irregularities on surfaces. |
| **4Pi** | Image resolution of <100 nm has been achieved. Imaging algorithms have been developed to enhance image resolution. A wide range of flourophores have been successfully imaged. The technique enables 3D images to be constructed. | In order to achieve good image quality complex experimental set-ups involving co-focalization and multi-photon excitation are required. Samples are required to be ultrathin on the order of 200 nm. Long accumulation times are also required. |
| **STED** | Image resolution of <30 nm has been achieved. The technique has been applied to a wide range of flourophores. A range of biosystems have been imaged using this technique. Relatively inexpensive experimental set-up based on diode lasers. The use of multi-colour imaging has been achieved. | Confocalization is required for the highest spatial resolution leading to long image accumulation times. Photobleaching of fluorophores is a problem that limits resolution. Experimental challenges creating the required optical transverse modes and pulse are present. |
| **SSIM** | Unlimited image resolution is possible using SSIM. Wide field imaging leading to fast image accumulation times. Relatively in-expensive experimental set-up. | A requirement of this technique however is that the sample be motionless. While SSIM has not been applied to image biosystems. Image reconstruction is required. Only one study of biosystems using SSIM has been reported. |
| **FIONA** | Position of the fluorophores emission can be determined with extremely high accuracly i.e. 1.5 nm. Wide field imaging leading to fast image accumulation times. | Flourophores are required to be diluted for imaging significantly reducing its potential for routine bio-imaging. Orientation of the fluorophore molecule can induce a |
| | | significant error when imaging with this method. |
| **PAINT** | High spatial resolution of c.a. 25 nm has been achieved. Images can be accumulated quickly with micrometer-sized images were recorded in seconds. A wide selection of probes may be applicable for PAINT imaging. Labelling with flourophores is not required. | A liquid interface was utilised in this technique. Mathematically estimation of individual flourophores position is required. Careful choice of flourophore is important to enable selectivity in imaging specific objects and systems. |
| **PALM** | Relatively simple experimental set-up based on a TIRF microscope. This approach has reported to enable the localized of molecules within these sections with an accuracy of 10 nm. | Special photoactivable flourophores are required. Flourophore stability limits imaging. Very long image accumulation times. Thin sample sections of 50 to 80 nm in height are imaged. |
| **SSL** | Theoretically unlimited resolution can be achieved. Potentially inexpensive experimental method. Either confocal or wide-field is possible with this approach. | The lenses are difficult to obtain as are not yet commercially available. Have not been applied to image biosystems using this experimental approach. |

Tab. 2. List of advantages and disadvantages of each ultrahigh resolution fluorescence imaging technique. * SWFM is standing wave fluorescence microscopy, 4Pi microscopy, STED is stimulated emission depletion fluorescence microscopy, PAINT is point accumulation for imaging in nanoscale topography, PALM is Photo-activated localization microscopy, SSL is super and/or hyper lense based techniques.